\documentclass[a4paper,12pt]{article}

\usepackage{amssymb,amsmath,cite}
\usepackage{graphicx}

\newcommand{\afn}[1]{\mathcal{E}_{#1}}         % <   |
\newcommand{\abs}{\bar{\alpha}_s}         % BFKL strong coupling
\newcommand{\as}{\alpha_s}                % strong coupling
\newcommand{\bs}{\boldsymbol}
\newcommand{\cut}{\mathrm{cut}}            % preclustering cut parameter
\newcommand{\Del}{\boldsymbol{\Delta}}          % delta mass-shell
\newcommand{\dif}{\mathrm{d}}             % finite-dimensional differential
\newcommand{\e}{\epsilon}
\newcommand{\esp}[1]{\mathrm{e}^{#1}}    % esponential
\newcommand{\gev}{\mathrm{\,GeV}}          % GeV
\newcommand{\imp}{\Longrightarrow}
\newcommand{\intC}{\boldsymbol{C}}         % integrated C_m
\newcommand{\intI}{\boldsymbol{\Psi}}         % integrated impact factor
\newcommand{\kt}{\boldsymbol{k}}
\newcommand{\Lqcd}{\Lambda_{\mathrm{QCD}}}  % Lambda_qcd
\newcommand{\ord}[1]{\mathcal{O}\left(#1\right)}
\newcommand{\nlo}{{\mathrm{NLO}}}          % NLO
\newcommand{\R}{\mathbb{R}}              % real
\newcommand{\ui}{\mathrm{i}}             % imaginary unit
\newcommand{\Z}{\mathbb{Z}}              % integers

\title{{\bf The NLO jet vertex\\ in the small-cone approximation\\ for kt and
  cone algorithms}}

\author{
   D.~Colferai and A.~Niccoli\\
   {\sl\small Dipartimento di Fisica e Astronomia, Universit\`a di Firenze and}\\
   {\sl\small INFN, Sezione di Firenze 50019 Sesto Fiorentino, Italy}
}

\date{}

\begin{document}
%@@@@@@@@@@@@@@@@@@@@@@@@@@@@@@@@@@@@@@@@@@@@@@@@@@@@@@@@@@@@@@@@@@@@@@@@@@@@@@@

\maketitle

\begin{abstract}
We determine the jet vertex for Mueller-Navelet jets and forward jets in the
small-cone approximation for two particular choices of jet algoritms: the kt
algorithm and the cone algorithm. These choices are motivated by the extensive
use of such algorithms in the phenomenology of jets. The differences with the
original calculations of the small-cone jet vertex by Ivanov and Papa, which
is found to be equivalent to a formerly algorithm proposed by Furman, are
shown at both analytic and numerical level, and turn out to be sizeable.
A detailed numerical study of the error introduced by the small-cone
approximation is also presented, for various observables of phenomenological
interest. For values of the jet ``radius'' $R=0.5$, the use of the small-cone
approximation amounts to an error of about 5\% at the level of cross section,
while it reduces to less than 2\% for ratios of distributions such as those
involved in the measure of the azimuthal decorrelation of dijets.
\end{abstract}

\vskip 1cm

%\begin{minipage}{0.9\textwidth}
%\begin{flushright}
%  Draft $ $Revision$ $ \\
%  $ $Date$ $
%\end{flushright}
%\end{minipage}
%\vskip 1cm

%%%%%%%%%%%%%%%%%%%%%%%%%%%%%%%%%%%%%%%%%%%%%%%%%%%%%%%%%%%%%%%%%%%%%%%%%%%%%%%%%
\section{Introduction\label{s:intro}}
%%%%%%%%%%%%%%%%%%%%%%%%%%%%%%%%%%%%%%%%%%%%%%%%%%%%%%%%%%%%%%%%%%%%%%%%%%%%%%%%%

The identification of high-energy (Regge) dynamics in QCD processes has been
long since theoretically investigated. With the advent of high-energy colliders
like HERA, Tevatron and LHC such studies have become possible at experimental
level too. The processes which are expected to be more sensitive to this
peculiar dynamical regime, where the center-of-mass energy $\sqrt{s}$ is much
larger than all hard scales involved in the scattering, are the so-called {\em
  Mueller-Navelet} (MN) {\em jets} at hadron-hadron colliders~\cite{MuNa87} and
{\em forward jets} at electron-hadron colliders~\cite{Mue91}. Both of them are
defined by the presence of QCD jets at large rapidity, accompanied by any
hadronic activity which is inclusively collected in the central region.

Such processes can be theoretically described by factorization formulae which
involve several ingredients: the partonic distribution functions (PDFs) of the
incoming hadron(s), the gluon Green's function (GGF) describing the high-energy
dynamics of emitted and exchanged partons --- mostly (reggeized) gluons --- and
finally the so-called jet vertices, describing the production of a forward jet
from the interaction of one incoming parton and a reggeized gluon. In the case
of an incoming electron, an additional quantity, the photon impact factor, has
to be considered too.

At present, all ingredients are known at next-to-leading level in the respective
parameters: the PDFs which resum logarithms of collinear type, the GGF resumming
logarithms of the energy, and the jet vertices (and impact-factors) which are
computed at finite perturbative order.

Focusing for definiteness on MN jets in hadron-hadron collision, the
process-dependent part of the cross section is represented by the jet
vertex~\cite{BaCoVa02}, which depends on the jet variables and the actual jet
algorithm.  By following the work of Ivanov and Papa~\cite{IP}, in this paper we
reconsider the computation of the jet vertex in the small-cone approximation
(SCA), namely for jets whose extension in the rapidity-azimuthal angle
$(y,\phi)$-plane is small.  In particular we apply their method to determine the
analytic expressions of the jet vertex for two particular choices of jet
algorithms: the {\em kt algorithm}%
\footnote{Here {\em kt algorithm} denotes the whole class of clustering
  algorithms based on ref.~\cite{CDSW93}, which may differ in the details of the
  recombination scheme and of the resolution variable, like the {\em anti-kt}
  and the {\em Cambrigde/Aachen} versions.}
\cite{CDSW93} and the {\em cone algorithm}~\cite{EKS89}. These are the mostly
used algorithms in modern jet phenomenology, in particular the kt one. On the
contrary, the algorithm used in~\cite{IP} can be traced back to the one
considered by Furman~\cite{Fur82} in early studies of QCD radiation, but not
used for practical purposes anymore, being infra-red unsafe.

The aim of our work is twofold: on one hand we want to give a precise estimate
of the error introduced by the small-cone approximation in the description of
QCD observables at high energies, i.e., at large rapidities, so as to possibly
justify its use in phenomenological analyses.  On the other hand, we want to
estimate the differences occurring by choosing different jet algorithms for the
same process.
%To our knowledge, up to now the latter issue has not been raised in studies on
%Mueller-Navelet jets, and our impression is that people believe the small-cone
%approximation to yield the same results for the various jet algorithms such as
%the three mentioned before.
The jet algorithm dependence has already been studied in the past, and a
detailed analysis of the cone and kt algorithms in the SCA was presented in
ref.~\cite{MuVo12} in the context of collinear factorization. Here we carry out
a similar analysis in the framework of high-energy (kt-dependent) factorization,
the basic tool for the description of the Regge regime in perturbative QCD.

One should keep in mind that the small-cone expressions are fully analytic
(before their convolution with the PDFs) and compact, and allow a simple
implementation in numerical codes that run much faster than those with the exact
jet vertices. For this reason the SCA jet vertices have already been
used\cite{CIMP13,CMSS13} in quantitative comparison with available
data\cite{cmsMNjet}. However, the experimental results were extracted by
clustering jets with the kt algorithm, while the SCA jet vertices used for the
theoretical calculation were those obtained with the Furman algorithm by Ivanov
and Papa (FIP).  Also the small-cone analysis performed in ref.~\cite{DSW13}
compared calculations with the exact jet vertices in the kt algorithm versus the
small-cone ones in the FIP algorithm.

Starting from these premises, and after reviewing in sec.~\ref{s:ts} the
theoretical setup for the description of Mueller-Navelet jets in terms of the
collinear and high-energy factorization formulae, in sec.~\ref{s:jetalg} we
discuss in detail the differences among the three (kt, cone and FIP) jet
algorithms in the relevant case of two near particles and we determine the
kinematical configurations in the limit of small jet ``radius'' $R$.

We then derive the small-cone jet vertices for the kt and cone algorithms in
sec.~\ref{s:scjv}, by computing their differences with respect to the FIP
algorithm induced by the different kinematical conditions.

In sec.~\ref{s:nc} we perform a numerical study in order to asses the
quantitative difference between the exact and small-cone jet vertex in the kt
algorithm, by comparing the vertices themselves as well as a typical
differential cross section for MN jets and some angular coefficients measuring
the azimuthal decorrelation between the jets. In addition we also determine the
discrepancies of the same quantities induced by a different choice of the jet
algorithm.

We discuss the results in sec.~\ref{s:conc}, where we conclude that the wrong
choice of algorithm causes sizeable errors on the predictions, while the SCA
within the same algorithm provides a good approximation to the exact quantities
and can therefore be used as a valuable tool for a quantitative description of
MN and forward jets.

%%%%%%%%%%%%%%%%%%%%%%%%%%%%%%%%%%%%%%%%%%%%%%%%%%%%%%%%%%%%%%%%%%%%%%%%%%%%%%%%%
\section{Theoretical setup\label{s:ts}}
%%%%%%%%%%%%%%%%%%%%%%%%%%%%%%%%%%%%%%%%%%%%%%%%%%%%%%%%%%%%%%%%%%%%%%%%%%%%%%%%%

%================================================================================
\subsection{Factorization\label{s:fact}}
%================================================================================

The process we are considering was suggested long ago by Mueller and
Navelet~\cite{MuNa87} in order to study the high-energy behaviour of QCD. It is
generated by the collision of two hadrons $H_{A,B}$ --- typically (anti)protons --- and is
characterized by the detection in the final state of two hard jets $J_{1,2}$ with
large rapidity separation:
\begin{equation}\label{proc}
  H_A + H_B \to J_1 + J_2 + X
\end{equation}
where $X$ represents any additional emission. Each jet
$J_i$ represents a cluster of particles grouped together according to some given jet
algorithm and is described by 3 variables: the rapidity $y_i$, the {\em transverse}
energy $E_i\equiv|\kt_{J,i}|$ and the azimuthal angle $\phi_i\equiv \arg(\kt_{J,i})$,
$\kt_{J,i}$ being the $i$-th jet transverse momentum.

The kinematical region where one expects the high-energy QCD dynamics to play an
important role is given by
\begin{equation}\label{hekin}
  s \equiv (p_A+p_B)^2 \gg E_1^2 \sim E_2^2 \gg \Lqcd^2 \;, \qquad
  |Y| \equiv |y_1 - y_2| \gg 1 \;,
\end{equation}
where the condition of hard jets ($E_i^2\gg\Lqcd^2$) is imposed for the
applicability of perturbation theory.

\begin{figure}[t]
  \centering
  \includegraphics[width=0.4\linewidth]{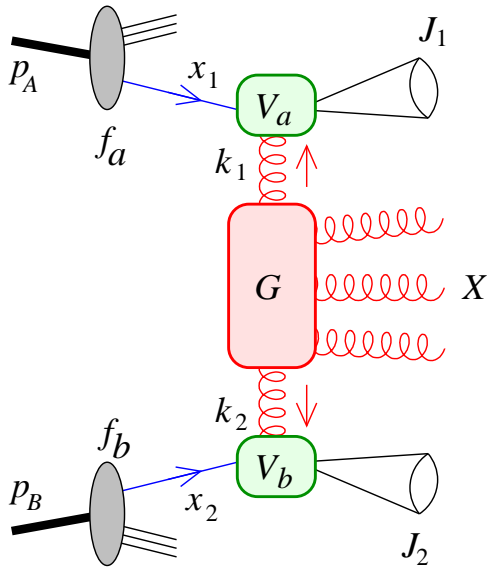}
  \caption{\it Diagrammatic representation of the collinear and high-energy
    factorization formula for Mueller-Navelet jet production: $f_{a,b}$
    represents the parton densities, $V_{a,b}$ the jet vertices and $G$ the
    gluon Green's function.}
  \label{f:fact}
\end{figure}

In the leading twist approximation, i.e., up to power suppressed correction in
the hard scale parameter $\Lqcd^2/E^2\ll1$, the hadronic cross section $\sigma$
can be factorized in the (longitudinal momentum fraction) convolution of two
partonic distribution functions (PDFs) $f_{a/H}(x)$ and a partonic cross section
$\hat\sigma$%~\cite{collFact}
\begin{equation}\label{collfact}
  \frac{\dif\sigma_{AB}(s)}{\dif J_1 \dif J_2} = \sum_{a,b}\int_0^1 \dif x_1 \dif x_2\;
  f_{a/A}(x_1) f_{b/B}(x_2) \frac{\dif\hat\sigma_{ab}(x_1 x_2 s)}{\dif J_1 \, \dif J_2}
  \;, \qquad \dif J_i \equiv \dif y_i \,\dif E_i \,\dif\phi_i \;,
\end{equation}
where $a,b\in\{q,g\}$ denote the parton flavours (quark or gluon) and $x_i$ the partonic
momentum fractions w.r.t.\ their parent hadrons.
In turn, in the high-energy Regge regime we are considering, the partonic cross
section for jet production can be factorized in a (transverse momentum)
convolution of (process dependent) jet vertices $V$ and a universal factor $G$
called gluon Green's function (GGF)%~\cite{heFact}
\begin{equation}\label{hefact}
  \frac{\dif\hat\sigma_{ab}(x_1 x_2 s)}{\dif J_1 \, \dif J_2} =
  \int\dif^2\kt_1\,\dif^2\kt_2 \; V_a(x_1,\kt_1;J_1) G(x_1 x_2 s, \kt_1, \kt_2)
  V_b(x_2,\kt_2;J_2) \;,
\end{equation}
where $\kt_i$ denotes the (reggeized) gluon transverse momentum flowing out from
the GGF and entering the jet vertex, while $\hat{s}\equiv x_1 x_2 s$ is the
center-of-mass energy squared of the partonic subsystem. The overall
factorization structure is depicted in fig.~\ref{f:fact}.

The PDFs are non-perturbative objects that depend also on the renormalization
and factorization scales $\mu_R$, $\mu_F$. While $\mu_R$ is introduced in the
renormalization of UV divergencies, $\mu_F$ enters in the treatment of the IR
collinear divergencies which are absorbed by the PDFs. The $\mu_F$ dependence of
the PDFs is governed by the DGLAP equations~\cite{DGLAP}, and their evolution
kernels (splitting functions) are known at next-to-next-to-leading order (NNLO).
By means of global fits they have been determined in a wide range of the
$(x,\mu_F^2)$ plane.

The GGF is the central object in high-energy QCD, in that it resums the
$\log(s)$ to all orders in perturbation theory. It obeys the BFKL
equation~\cite{BFKL}
\begin{align}\label{bfkl}
  \omega G_{\omega}(\kt_1,\kt_2) &= \delta^2(\kt_1+\kt_2) + \int\dif^2\kt\;
  K(\kt_1,\kt) G_{\omega}(\kt,\kt_2) \\
  G(\hat{s},\kt_1,\kt_2) &=
  \int_{-\ui\infty}^{\ui\infty}\frac{\dif\omega}{2\pi\ui} \;
  \left(\frac{\hat{s}}{s_0}\right)^\omega G_{\omega}(\kt_1,\kt_2) \;,
  \label{Gomega}
\end{align}
where $G_\omega$ is the Mellin transform of $G(\hat{s})$, defined in terms of an
arbitrary energy scale $s_0$. The BFKL kernel $K=\as K^{(0)} + \as^2 K^{(1)}$ is
known in next-to-leading-logarithmic (NLL) approximation. The coefficient
$K^{(1)}$ depends on the choice of both $\mu_R$ and $s_0$.

Finally, the jet vertices are perturbative finite objects without energy
($\sqrt{s}$) dependence, and are known in NLO approximation:
$V=\as V^{(0)}+\as^2 V^{(1)}$. They depend on the jet variables $y,E,\phi$ and
also on the arbitrary scales $\mu_R,\mu_F,s_0$, in such a way that the hadronic
cross section be independent of those scales up to NLL terms, i.e., the scale
dependence is present only in the terms of relative order $\as^2(\as\log(s))^n$.

It is apparent that the determination of the Mueller-Navelet (MN) jet cross
section involves quite a number of integrals, both in the factorization
formulae~(\ref{collfact},\ref{hefact}), and in the determination of $G$ and
$V$, as can be checked from their explicit expressions~\cite{BaCoVa02}.
Furthermore, when comparing the theoretical predictions with experiments, one
needs integrated cross sections in some of the jet variables, in order to comply
with the experimental binning. Such integrations are mostly done numerically,
and this would require a large amount of computing resources or time for reaching
a precision at the level of 1\%.

In order to cope with such a problem, two techniques can be exploited so as to
reduce the computing time and improve convergence:
\begin{itemize}
\item to project the GGF and jet vertices on a complete set of functions which
  respects the symmetries of the process (e.g., azimuthal invariance);
\item to use an approximated and simpler version of the jet vertices.
\end{itemize}
These methods are often used in BFKL phenomenological analyses, and we shall
illustrate them in the following subsections: the former in order to set up the
theoretical framework and the main notations; the latter in order to introduce
the main subject of this paper.

%================================================================================
\subsection{Representation in Mellin space\label{s:rms}}
%================================================================================

The first method to reduce computing time does not involve any approximation, at
least at the NLL level of accuracy we are working with. It is better illustrated
in the LL approximation, where the strong coupling $\as$ is fixed and thus the
BFKL kernel and jet vertices, in addition of being invariant under azimuthal
rotations, are also scale invariant.%
\footnote{Strictly speaking, in our notations they are homogeneous functions of
  the transverse momenta $\kt$, $\kt_J$.}  In this case, a Fourier-Mellin
transform diagonalizes the transverse integrations and the ensuing expressions
are considerably simpler to evaluate.

One can proceed in this way: first of all, let's exploit the fact that, in the
LL approximation, each partonic momentum fraction coincides with the
corresponding jet's longitudinal momentum fraction (because of a $\delta(x-x_J)$
in $V^{(0)}$):
\begin{equation}\label{xJ}
  x = x_J \equiv E\esp{\pm y}/\sqrt{s} \qquad
  (+,- \text{ for jet } 1,2) \;,
\end{equation}
so that $\hat{s} = x_1 x_2 s = \esp{Y} E_1 E_2$, where $Y\equiv|y_1-y_2|$ is the
rapidity distance between the jets.  If we adopt the convenient and natural
choice of $s_0=E_1 E_2$ as energy scale, the Green's function~(\ref{Gomega}) is
now independent of the partonic momentum fractions:
\begin{equation}\label{GY}
  G(\hat{s},\kt_1,\kt_2) =
  \int_{-\ui\infty}^{\ui\infty}\frac{\dif\omega}{2\pi\ui} \;
  \esp{\omega Y} G_{\omega}(\kt_1,\kt_2) \equiv G(Y,\kt_1,\kt_2)
  \;,\qquad (s_0 = E_1 E_2)
\end{equation}

Secondly, we introduce the {\em impact factor} by integrating a jet vertex with
the corresponding parton density:
\begin{equation}\label{impf}
  \Phi_A(\kt;J) \equiv \sum_a\int_0^1\dif x\; f_{a/A}(x) V_a(x,\kt;J) \;.
\end{equation}
so that we can rewrite the factorization formula in the form
\begin{equation}\label{sigmaFact}
  \frac{\dif\sigma_{AB}}{\dif J_1\,\dif J_2} =
  \int\dif^2\kt_1\,\dif^2\kt_2\; \Phi_A(\kt_1) G(Y,\kt_1,\kt_2) \Phi_B(\kt_2) \;.
\end{equation}

At this point we project kernel and vertices onto the eigenfunctions of the LL
BFKL kernel
\begin{equation}\label{En}
  \afn{n\nu}(\kt) \equiv \frac1{\sqrt{2}\pi} |\kt^2|^{\ui\nu-\frac12}
  \esp{\ui n\phi} \;, \qquad(n\in\Z,\;\nu\in\R) \;,
\end{equation}
satisfying the completeness relation
\begin{equation}\label{comp}
  \sum_{n\in\Z}\int_{-\infty}^{\infty} \dif\nu\;
  \afn{n\nu}(\kt) \afn{n\nu}^*(\kt') = \delta^2(\kt-\kt')
\end{equation}
and providing the LL eigenvalue function $\chi^{(0)}_{n\nu}$
\begin{align}
  [K^{(0)} \afn{n\nu}](\kt) &\equiv \int\dif^2\kt'\; K^{(0)}(\kt,\kt')
  \afn{n\nu}(\kt') = \chi^{(0)}_{n\nu} \afn{n\nu}(\kt) \label{K0E}\\
  \chi^{(0)}_{n\nu} &= 2\psi(1)-\psi\left(\frac{1+n}{2}+\ui\nu\right)
  -\psi\left(\frac{1+n}{2}-\ui\nu\right) \;.
  \label{chi0}
\end{align}

Finally, by inserting a completeness~(\ref{comp}) between each pair of factors
in~(\ref{sigmaFact}), we arrive at the convenient expression for the
differential cross section
\begin{equation}\label{sigmaMel}
  \frac{\dif\sigma_{AB}}{\dif J_1\,\dif J_2} =
  \sum_n (-1)^n \int\dif\nu\;\Phi_{A\,n\nu} \; G_{n\nu}(Y) \; \Phi_{B\,n\nu}^*\;,
\end{equation}
where in the last equality we have used the Fourier-Mellin transforms
\begin{align}\label{Phin}
  \int\dif^2\kt\;\Phi(\kt;J) \afn{n\nu}(\kt) &\equiv \Phi_{n\nu}(J) \\
  \int\dif^2\kt\,\dif^2\kt'\; \afn{n\nu}^*(\kt)G(Y,\kt,\kt') \afn{n'\nu'}(\kt')
  &\equiv G_{n\nu}(Y) (-1)^n \delta_{nn'}\delta(\nu-\nu') \;. \label{Gn}
\end{align}
The delta functions on the r.h.s.\ of eq.~(\ref{Gn}) are just a consequence of
the azimuthal- and scale-invariance of the kernel, and allow us to trade two
bidimensional integrals for a sum and a simple integral.

The azimuthal correlation of the MN jets is usually measured by means of the
Fourier coefficients ($m\in\Z$)
\begin{equation}\label{defCm}
  C_m(E_1,y_1;E_2,y_2) \equiv
  \int_0^{2\pi}\dif\phi_1\dif\phi_2 \;\cos\big(m(\phi_1-\phi_2-\pi)\big)
  \frac{\dif\sigma}{\dif J_1\,\dif J_2} \;.
\end{equation}
Because of the azimuthal- and scale-invariance of the vertices, it is easy to
show that
\begin{equation}\label{phiStruct}
  \Phi_{n\nu}(y,E,\phi) = \esp{\ui n \phi} (E^2)^{\ui\nu-1}
  \Psi_{n\nu}(x_J) \;, \qquad \Psi_{n\nu} = \Psi_{-n\nu}
\end{equation}
where $\Psi$ is dimensionless,%
\footnote{We shall sometimes refer to the reduced impact factor $\Psi$ as jet
  vertex.}
thus obtaining a factorization formula with just one integration:
\begin{equation}\label{Cm1}
  C_m = \left(\frac{2\pi}{E_1 E_2}\right)^2 \int\dif\nu
  \left(\frac{E_1^2}{E_2^2}\right)^{\ui\nu} \Psi_{A\,m\nu} \; G_{m\nu}(Y) \;
  \Psi_{B\,m\nu}^* \;.
\end{equation}

Such a structure is preserved in the NLL approximation too, provided the impact
factors and GGF are suitably modified in order to take into account the loss of
scale-invariance due to the renormalization procedure and to the factorization
of collinear singularities, which translates in a dependence on $\log(E/\mu_R)$,
$\log(E/\mu_F)$ and $\log(E/\sqrt{s_0})$ of the GGF and impact factors.  The
expression of the GGF in Mellin-space is very simply expressed in terms of the
eigenvalue $\chi_{n\nu}$ of the BFKL kernel:
\begin{equation}\label{Gnnu}
  G_{n\nu}(Y) = \esp{Y\chi_{n\nu}} \;, \qquad
  \chi_{n\nu} = \abs \chi_{n\nu}^{(0)} + \abs^2 \chi_{n\nu}^{(1)}
\end{equation}
However, the computation of $\Psi_{n\nu}$ at NLL level involves several
integrations. The main advantage of this procedure is that such integrations can
be done once and for all for each set of one-jet variables. Nevertheless, such
computations can still be rather lengthy, and the use of an approximate
expression of the impact factors --- to be described in the next subsection ---
turns out to be very convenient.

Let us conclude this section by noticing that the factorization
formula~(\ref{Cm1}) is very useful also in the case of cross-section integrated
in jet energies. In fact, a double integral in $E_1$ and $E_2$ factorizes into
the product of simple integrals of the impact factors (provided the integration
domain $D=I_1\times I_2$ can be factorized into the cartesian product of two
one-dimensional sets)
\begin{align}\label{intCm}
%  \intC_m(y_1,y_2) &\equiv \int_{E_1^{\min}}^{E_1^{\max}}\dif E_1
%  \int_{E_2^{\min}}^{E_2^{\max}} \dif E_2 \; C_m \\ \nonumber
  \intC_m(y_1,y_2) &\equiv \int_{D=I_1\times I_2}\dif E_1 \dif E_2 \; C_m
  \\ \nonumber
  &= (2\pi)^2 \int\dif\nu \; G_{m\nu}(Y)
  \left[\int_{I_1} \dif E_1 \; (E_1^2)^{\ui\nu-1} \Psi_{A\,m\nu} \right] \;
  \left[\int_{I_2} \dif E_2 \; (E_2^2)^{\ui\nu-1} \Psi_{B\,m\nu} \right]^*
  \\ \nonumber
  &\equiv (2\pi)^2 \int\dif\nu \; G_{m\nu}(Y) \; \intI_{A\,m\nu}(y_1) \;
  \intI_{B\,m\nu}^*(y_2) \;.
\end{align}
In this case, the integrated impact factors $\intI$ can be computed
independently and stored in suitable grids, thus reducing a lot the
computational effort of the phase-space integration. In the general case
$D\neq I_1\times I_2$, however, the expression~(\ref{Cm1}) has to be numerically
integrated in energy (and possibly in rapidity), and a suitable approximation
(like the SCA) could be a very valuable tool to diminish the computing demand.

%================================================================================
\subsection{Small-cone approximation\label{s:sca}}
%================================================================================

In order to study the behaviour of the jet vertex for small values of the
``radius'' $R$ --- to be precisely defined later on --- and possibly to speed up
the computation of the jet impact factor $\Psi_{n\nu}$ and of its
energy-integrated version $\intI_{n\nu}$, one can use the small-cone
approximation (SCA), as suggested and derived in~\cite{IP}. The dependence of
the impact factor on the jet radius $R$ has the form~\cite{MuVo12} ($\mu$ is a
shorthand for $\mu_R$, $\mu_R$ and $\sqrt{s_0}$)
\begin{equation}\label{Rdep}
  \Psi_{n\nu}\Big(x_J,\log\frac{E}{\mu};R\Big) = A_{\nu}(x_J) \log(R)
  + B_{n\nu}\Big(x_J,\log\frac{E}{\mu}\Big) + \ord{R^2}
\end{equation}
and the analytic expressions for the coefficients $A,B$ were explicitly
computed~\cite{IP} for a particular jet algorithm (FIP).

However, whereas the coefficient $A$ of $\log(R)$ depends only on the incoming
hadron and is given in terms of the usual splitting functions as
\begin{align}\label{A}
 A_{\nu} = -\as^2\frac{\sqrt{N_c^2-1}}{\sqrt{2}\pi^2 N_c} x_J
 \int_{x_J}^1\dif\zeta\; \zeta^{-2\ui\nu}
 & \bigg\{ \big[P_{qq}(\zeta)+P_{gq}(\zeta)\big]
 \sum_{a\in\{q,\bar{q}\}} f_{a}\Big(\frac{x_J}{\zeta}\Big) \\
 & + \big[P_{gg}(\zeta)+2 n_f P_{qg}(\zeta)\big]
 f_{g}\Big(\frac{x_J}{\zeta}\Big) \bigg\} \;,
\end{align}
the constant term $B$ depends also on the details of the jet algorithm. Ivanov
and Papa~\cite{IP} computed such coefficient for an algorithm which was used in
pioneering work on QCD jets by Furman~\cite{Fur82} --- we shall refer to it as
FIP algorithm --- which, however, is no more used in present day phenomenology.

The main purpose of our paper is to derive such coefficient for the two mostly
used algorithms of QCD analysis, namely the cone-algorithm and the kt-algorithm.
The computations can be repeated by following the procedure of~\cite{IP}.

The expressions for the jet vertices at LL and NLL level are extracted from the
perturbative calculation of processes with two incoming partons producing 2-jet
at LO and NLO respectively. At LO the amplitudes have just two partons in the
final states, each of which is identified with a jet. The jets are emitted
back-to-back in the azimuthal direction and have no substructure. Therefore no
dependence on the algorithm is found at LO.

The same is true at NLO as far as the virtual corrections are concerned. On the
other hand, the NLO real corrections involve 3 partons in the final state,
therefore a jet can be constituted by either one or two of them. In the case of
1-parton (simple) jet, all algorithms are designed in such a way that no further
emission is found within a region of radius $R$ in the $(y,\phi)$ plane around
the position of that parton.

Therefore, the differences among the algorithms are to be found in the 2-parton
(composite) jet configurations. In the following section we will carefully
compare the definitions the {\it cone}, {\it kt}, and {\it FIP} jet
algorithms and, from their differences, we shall compute the small-cone impact
factors for the cone and kt jets.

%%%%%%%%%%%%%%%%%%%%%%%%%%%%%%%%%%%%%%%%%%%%%%%%%%%%%%%%%%%%%%%%%%%%%%%%%%%%%%%%
\section{Jet algorithms\label{s:jetalg}}
%%%%%%%%%%%%%%%%%%%%%%%%%%%%%%%%%%%%%%%%%%%%%%%%%%%%%%%%%%%%%%%%%%%%%%%%%%%%%%%%

%================================================================================
\subsection{The cone algorithm\label{s:cone}}
%================================================================================

According to ref.~\cite{EKS89}, when two partons $p_1$ and $p_2$ are combined
into one jet of radius $R$, the resulting jet variables $(y,E,\phi)$ are defined
to be
\begin{equation}\label{conejet}
  E = E_1 + E_2 \;, \qquad y = \frac{y_1 E_1 + y_2 E_2}{E} \;, \qquad
  \phi = \frac{\phi_1 E_1 + \phi_2 E_2}{E} \;.
\end{equation}
To determine whether the two partons are to be combined, we see if they fit in a
cone of radius $R$ about the jet axis in the $(y,\phi)$ plane. In practice, by
denoting with $\Omega_{ij}$ the $(y,\phi)$-distance
\begin{equation}\label{dij}
  \Omega_{ij}^2 \equiv (y_i-y_j)^2 + (\phi_i-\phi_j)^2
\end{equation}
one requires $\Omega_{1J}<R$ and $\Omega_{2J}<R$ for the composite jet, which
amounts the condition
\begin{equation}\label{coneComp}
  \Omega_{12} < R \frac{E_1+E_2}{\max(E_1,E_2)} \;.
\end{equation}
A simple jet can then be defined only if the two partons cannot be combined,
i.e., provided
\begin{equation}\label{coneSimple}
  \Omega_{12} > R \frac{E_1+E_2}{\max(E_1,E_2)} \;.
\end{equation}

%================================================================================
\subsection{The kt algorithm\label{s:kta}}
%================================================================================

According to ref.~\cite{CDSW93}, the kt clustering algoritm consists in an
iterative procedure which is based on comparing a set of resolution variables of
single-particle $d_{iB}$ (B for beam) and pairs $d_{ij}$
\begin{equation}\label{resVar}
  d_{iB}\equiv E_i^2 \;, \qquad
  d_{ij} \equiv \min(E_i^2,E_j^2)\, \frac{\Omega_{ij}^2}{R^2} \;.
\end{equation}
One then considers the smallest one: if it is a $d_{iB}$, particle $i$ is thrown
in the beam basket and removed from the list; if it is a $d_{ij}$ then particles
$i$ and $j$ are merged into a pseudoparticle $\{ij\}$ --- e.g., by using the
recombination scheme~(\ref{conejet}). The procedure is repeated from the
beginning, until all resolution variables are greater than some (hard) stopping
parameter $d_{\cut} \gg \Lqcd^2$.

In the case of three partons in the final state there are these possibilities:
\begin{itemize}
\item One of the $d_{iB}$, say $d_{1B}$ is the smallest resolution variable.
  \begin{itemize}
  \item If $d_{1B} < d_{\cut}$ then particle 1 belongs to the beam and we are
    left with two partons which, if not in the beam, form two simple jets;
  \item If $d_{1B} > d_{\cut}$ then the clustering stops and all three particles
    form simple jets;
  \end{itemize}
\item One of the $d_{ij}$, say $d_{12}$ is the smallest resolution variable. In
  this case 1 and 2 are merged into a pseudoparticle \{12\} and one considers
  three resolution variables: $d_{\{12\}B}$, $d_{3B}$ and $d_{\{12\}3}$.
    \begin{itemize}
    \item If the smallest is larger than $d_{\cut}$ clustering stops and we have one
      simple jet \{3\} and one composite jet \{12\};
    \item If the smallest is less than $d_{\cut}$, a (pseudo)particle
      belongs to the beam and we cannot have two jets in the final state;
    \end{itemize}
\end{itemize}
To summarize, we have a composite jet, say \{12\}, only if
\begin{equation}\label{ktComp}
  d_{12} < d_{iB} \;\forall i \qquad\imp\qquad \Omega_{12} < R \;.
\end{equation}
In the other case
\begin{equation}\label{ktSimple}
  \Omega_{12} > R
\end{equation}
only simple jets are present.

%================================================================================
\subsection{The Furman algorithm\label{s:ipa}}
%================================================================================

In~\cite{IP} Ivanov and Papa define a jet as a set of particles within a
cone of radius $R$. To be more precise, they require that:
\begin{itemize}
\item the jet's momentum is the sum of the particles' momenta;
\item all and only the particles of the jet belong to a circle of radius $R$ in the
$(y,\phi)$ plane and centered at the jet's momentum.
\end{itemize}
This is just the jet definition of Furman~\cite{Fur82} used in the '80s for
early phenomenology of NLO QCD jets --- we denote it ``FIP algorithm''.

In the case under study, where at most two particles (1 and 2) can form a jet,
the condition for a composite jet is nothing but the prescription adopted in the
cone algorithm: two particles such that
\begin{equation}\label{ipComp}
  \Omega_{12} < R \frac{E_1+E_2}{\max(E_1,E_2)}
\end{equation}
can be considered as a composite jet. On the other hand, particle 1 can form a
simple jet if no other particle is found within a distance $R$ from it, namely
\begin{equation}\label{ipSimple}
  \Omega_{12} > R \;.
\end{equation}
This definition is somewhat pathological, because it may happen that a given
configuration can give rise to both a composite jet and two simple jets. This
fact can be easily understood in the case of two particles with the same
transverse energy and whose distance satisfy $R < \Omega_{12} < 2R$. They can
form simple jets because a cone of radius $R$ centered on either particle does
not contain the other; on the other hand, a cone of radius $R$ centered halfway
the two particles contains both of them. For this reason, it is not possible to
extend such jet definition into an IR-safe algorithm to all orders, i.e., with
an arbitrary number of particles, hence it has been abandoned in favour of
better algorithms like the cone and especially the kt one.

In any respect, this definition is different from both the cone algorithm (in
the case of simple jets) and the kt algorithm (in the case of composite jets),
and yields different results when used to defined any observable. It is the main
purpose of this paper to derive the correct expressions for the jet vertices in
the SCA for the kt algorithm and also for the cone one.

%%%%%%%%%%%%%%%%%%%%%%%%%%%%%%%%%%%%%%%%%%%%%%%%%%%%%%%%%%%%%%%%%%%%%%%%%%%%%%%%
\section{Small cone jet vertices\label{s:scjv}}
%%%%%%%%%%%%%%%%%%%%%%%%%%%%%%%%%%%%%%%%%%%%%%%%%%%%%%%%%%%%%%%%%%%%%%%%%%%%%%%%

In this section we shall compute the SCA jet vertex in the cone and kt algoritm.
The idea is to identify and calculate the contributions differing from those of
the original paper~\cite{IP} where the FIP algorithm was adopted.

\begin{figure}[t]
  \centering
  \includegraphics[width=0.22\textwidth]{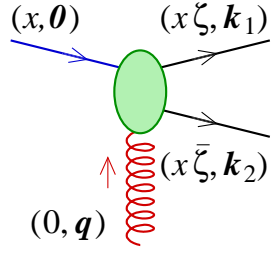}
  \caption{\it Kinematics of the fragmentation region of hadron A: the collision
    of the incoming parton (blue) and of the Regge gluon (red) produces a pair
    of outgoing partons (black). In parentheses the longitudinal momentum
    fraction and the transverse momentum of each particle.}
  \label{f:fragKinem}
\end{figure}

As explained in the previous section, for dijet production at NLO, the
differences among jet algorithms occur only in the way that two partons can be
combined to form a jet. In the BFKL approach, two partons (1 and 2) in the
fragmentation region of an incoming hadron are produced by the interaction of a
parton stemming from the hadron and a reggeized gluon, as depicted in
fig.~\ref{f:fragKinem}.

In the high-energy kinematics, the incoming parton has just a fraction $x$ of
the pure longitudinal momentum of the hadron, while the reggeized gluon's
momentum $q$ is essentially transverse. Following Ivanov-Papa (IP)~\cite{IP} we
work in dimensional regularization ($D=4+2\e$) and we indicate with $k_1$ ($k_2$)
the ($2+2\e$)-dimensional transverse momentum of the outgoing parton 1 (2), and
with $\zeta$ ($\bar\zeta\equiv 1-\zeta$) its longitudinal momentum fraction with
respect to the incoming parton.  In terms of these variables, the relative
rapidity and azimuthal angle between the two outgoing partons are
\begin{equation}\label{deltas}
  \Delta y = \frac12\log\frac{\zeta^2 k_2^2}{\bar\zeta^2 k_1^2} \;, \qquad
  \Delta\phi = \arccos\frac{k_1\cdot k_2}{|k_1|\,|k_2|} \;, \qquad \bar\zeta
  \equiv 1-\zeta \;.
\end{equation}

We shall present our results by using the notations of ref.~\cite{IP}, namely in
term of their ``$I$'' quantities, which are related to our definition of jet
impact factor~(\ref{Phin}) by
\begin{equation}\label{notation}
 \Phi_{n\nu}(y,E,\phi)
 = \as \frac{\sqrt{N_c^2-1}}{\sqrt{2}\pi N_c} \frac{x_J}{E} \, I(n,\nu;y,E,\phi)
  \;, \qquad I = \sum_{i,f} (I^R_{i;f} + I^V_{i;f}) \;,
\end{equation}
where $i$ ($f$) are labels for the initial (final) state of the sub-processes
contributing to the cross section, while the superscripts $R$ and $V$ denote
real and virtual parts respectively.

%================================================================================
\subsection{Vertex for the cone algorithm\label{s:cav}}
%================================================================================

The condition for composite jet in the cone algorithm~(\ref{coneComp}) coincides
with the one adopted by FIP~(\ref{ipComp}), thus no difference is expected for
the corresponding contributions.

In contrast, the conditions for simple jets~(\ref{coneSimple}) and
(\ref{ipSimple}) are different, causing different contributions to the
jet vertices.  By using the notations of ref.~\cite{IP}, $k=k_1$ is the
transverse momentum of the parton forming the simple jet, $q-k=k_2$ is the
transverse momentum of the parton outside the jet (also called ``spectator''),
and one introduces the transverse vector $\Del$
\begin{equation}\label{delta}
  q = \frac{k}{\zeta} + \Del
\end{equation}
which vanishes when the two partons are collinear. In fact, for small $\Del$,
we have
\begin{equation}\label{omegasc}
  \Omega_{12}^2 \equiv \Delta\phi_{12}^2 + \Delta y_{12}^2
  \simeq \frac{\zeta^2}{\bar\zeta^2}\frac{\Del^2}{k^2}
\end{equation}
and also
\begin{equation}\label{encomb}
  \frac{E_1+E_2}{\max(E_1,E_2)} = \frac{|k_1|+|k_2|}{\max(|k_1|,|k_2|)} \simeq
  \frac{|k|\left(1+\frac{\bar\zeta}{\zeta}\right)}{|k|\max\left(1,\frac{\bar\zeta}{\zeta}\right)}
  = \frac1{\max(\zeta,\bar\zeta)}
\end{equation}
Therefore, the simple-jet condition~(\ref{coneComp}) becomes
\begin{equation}\label{coneDelta}
  |\Del| > \frac{\bar\zeta}{\zeta} |k|\frac{R}{\max(\zeta,\bar\zeta)} \;,
\end{equation}
at variance with the FIP condition~(\ref{ipComp}) which, expressed in terms of
$\Del$, reads like eq.~(\ref{coneDelta}) but without the denominator
$\max(\zeta,\bar\zeta)$~\cite{IP}.

In order to identify the contributions of the FIP jet vertex that have to be
modified in the cone algorithm, let us recall that the simple jet configurations
were computed in two steps: {\it 1)} by allowing the spectator parton to span
the whole phase space and then {\it 2)} by subtracting the contribution stemming
from the spectator parton inside the jet cone. Therefore, we have to modify only
the subtractions by replacing $R\to R/\max(\zeta,\bar\zeta)$.

The {\bf quark initiated} jet vertex (quark+Regge-gluon $\to$ quark+gluon, see
fig.~\ref{f:fragKinem}) has two such subtractions, one for the quark-jet and one
for the gluon-jet, whose results are reported in eqs.~(5.36) and (5.38) of
ref.~\cite{IP} respectively. In the gluon-jet term the substitution $R\to
R/\max(\zeta,\bar\zeta)$ is straightforward:
\begin{align}
  I_{q;g,-q}^R &= (5.36)[1] \to
  -\frac{\as}{2\pi}\frac{\Gamma(1-\e)}{\e(4\pi)^e}\frac{\Gamma^2(1+\e)}{\Gamma(1+2\e)}
  (k^2)^{\gamma+\e-\frac{n}{2}} (k\cdot l)^n
  \int_{x_J}^1\frac{\dif\zeta}{\zeta}\;\zeta^{-2\gamma} \sum_{a=q,\bar{q}}
  f_a\left(\frac{x_J}{\zeta}\right) \nonumber \\
 &\qquad\qquad\times\left(\frac{R}{\max(\zeta,\bar\zeta)}\right)^{2\e}
 \left[P_{gq}(\zeta)
  \left(1+2\e\log\frac{\bar\zeta}{\zeta}\right)+\e C_F\zeta\right] \nonumber \\
 &= I_{q;g,-q}^R - \frac{\as}{2\pi}  (k^2)^{\gamma-\frac{n}{2}} (k\cdot l)^n
  \int_{x_J}^1\frac{\dif\zeta}{\zeta}\;\zeta^{-2\gamma} \sum_{a=q,\bar{q}}
  f_a\left(\frac{x_J}{\zeta}\right)
  P_{gq}(\zeta) 2\log\big(\max(\zeta,\bar\zeta)\big) \;,
\label{diffIqgq}
\end{align}
where $\gamma\equiv\ui\nu-\frac12$ and $l\equiv e_1+\ui e_2$ is a complex vector
lying only in the first two of the $2+2\e$ transverse dimensions.

In the quark-jet one has to proceed more carefully, because of the presence of a
double pole in $\e$ multiplying $(R/\max(\zeta,\bar\zeta))^{\e}$. This would
generate, among other things, modified simple poles and also finite double logs.
However, since the double pole is multiplied by a $\delta(1-\zeta)$, for these
terms $\max(\zeta,\bar\zeta)=1$ and the outcome is identical to the FIP
algorithm. The only difference comes from the simple pole in front of the
$P_{qq}$ splitting function, and we obtain
\begin{align}
  I_{q;q,-g}^R &= (5.38)[1] \nonumber \\
 &\to I_{q;q,-g}^R - \frac{\as}{2\pi}  (k^2)^{\gamma-\frac{n}{2}} (k\cdot l)^n
  \int_{x_J}^1\frac{\dif\zeta}{\zeta}\;\zeta^{-2\gamma} \sum_{a=q,\bar{q}}
  f_a\left(\frac{x_J}{\zeta}\right)
  P_{qq}(\zeta) 2\log\big(\max(\zeta,\bar\zeta)\big) \;.
  \label{diffIqqg}
\end{align}

The {\bf gluon initiated} jet vertex has two subtractions too, one for the
$q\bar{q}$ final state and one for the $gg$ final state, whose results are
reported in eqs.~(5.49)%
\footnote{We note a misprint in ref.~\cite{IP}: in eqs.~(5.48-49) the first
  subscript of $I^R$ should be $g$ instead of $q$. Also the last subscript in
  eq.~(5.38) should be $-g$ instead of $-q$.}
and (5.56) of ref.~\cite{IP} respectively. The situation
is very similar to that of the quark initiated vertex: the subtraction in the
(anti)quark jet contains a simple pole times the $P_{qg}(\zeta)$ splitting
function in front of the $R^{2\e}$ factor, while the gluon jet contains both
single and double poles, the former with the $P_{gg}(\zeta)$ splitting function
and the latter with the $\delta(1-\zeta)$ distribution.
For both types of jets the substitution $R\to R/\max(\zeta,\bar\zeta)$ simply
amounts to finite contributions proportional to
$\log\big(\max(\zeta,\bar\zeta)\big)$:
\begin{align}
  I_{g;q,-\bar{q}}^R &+ I_{g;\bar{q},-q}^R + I_{g;g,-g}^R
  = (5.49)[1] + \{ q \leftrightarrow \bar{q} \} + (5.56)[1] \nonumber \\
 &\to I_{g;q,-\bar{q}}^R + I_{g;\bar{q},-q}^R + I_{g;g,-g}^R   \label{diffIg} \\
 & - \frac{\as}{2\pi}  (k^2)^{\gamma-\frac{n}{2}} (k\cdot l)^n
  \int_{x_J}^1\frac{\dif\zeta}{\zeta}\;\zeta^{-2\gamma} \sum_{a=q,\bar{q}}
  f_a\left(\frac{x_J}{\zeta}\right)
  \left[2 n_f P_{qg}(\zeta) + P_{gg}(\zeta) \right]
  2\log\max(\zeta,\bar\zeta) \;. \nonumber
\end{align}

To sum up, the jet vertex for the cone algorithm in the small-cone approximation
is obtained by replacing $R\to R/\max(\zeta,\bar\zeta)$ in the final formulae
(5.39) and (5.57) of~\cite{IP}. The complete expressions are written in
eqs.~(\ref{finalQ},\ref{finalG}).

%================================================================================
\subsection{Vertex for the kt algorithm\label{s:kav}}
%================================================================================

The condition for simple jet in the kt algorithm~(\ref{ktSimple}) coincides with
the one adopted by FIP~(\ref{ipSimple}), while the conditions for composite
jet~(\ref{ktComp}) and (\ref{ipComp}) are different. In the composite jet
configuration the jet's transverse momentum is $k=k_1+k_2=q$ and it is
convenient to define the auxiliary transverse vector $\Del$ as
\begin{equation}\label{Del2}
  k_1 = \zeta k + \Del
\end{equation}
thus obtaining
\begin{equation}\label{omegacc}
  \Omega_{12}^2 = \frac{\Del^2}{k^2 \zeta^2 \bar{\zeta}^2}
\end{equation}
The composite jet condition~(\ref{ktComp}) becomes
\begin{equation}\label{ktDelta}
  |\Del| < \zeta\bar\zeta |k| R \;,
\end{equation}
at variance with the FIP condition~(\ref{ipComp}) which reads~\cite{IP}
$|\Del|<\min(\zeta,\bar\zeta) |k| R$.

The corresponding contributions of~\cite{IP}, that have to be modified in order
to recover the jet vertex in the kt algorithm, are found in sec.~5.1.1.c for the
quark initiated vertex and secs.~5.2.1.b and 5.2.2.b for the gluon initiated
one. In all such cases the modification amounts to replace
\begin{equation}\label{deltaMaxRep}
  |\Delta_{\max}| = \min(\zeta,\bar\zeta) |k| R \to \zeta\bar\zeta |k| R \;,
\end{equation}
and finally to substitute $\min(\zeta,\bar\zeta)\to\zeta\bar\zeta$ in the
$\zeta$-integral with the relevant splitting function.

Explicitly, in the {\bf quark-initiated} case, such integrals for the FIP and kt
algoritms reads respectively
\begin{align}\label{IqqgFIP}
  I_{q;q+g}^{R\;(\mathrm{FIP})} &\propto \int_0^1\dif\zeta\;
  [\min(\zeta,\bar\zeta)]^{2\e} \frac{1+\bar\zeta^2+\e\zeta^2}{\zeta}
  = \frac1{\e} -\frac32 +\left(\frac72 -\frac{\pi^2}{3}+3\log 2\right)\e \\
  I_{q;q+g}^{R\;(\mathrm{kt})} &\propto \int_0^1\dif\zeta\; \quad\;\;
  (\zeta\bar\zeta)^{2\e} \quad \frac{1+\bar\zeta^2+\e\zeta^2}{\zeta}
  = \frac1{\e} -\frac32 +\left(\frac{13}{2} -\frac{2\pi^2}{3}\right)\e \;.
  \label{IqqgKT}
\end{align}
Their difference $(3-\pi^2/3-3\log2)\e$, when multiplied by the overall $1/\e$ pole,
provides a finite contribution that has to be added to the IP result in order to
obtain the proper expression for the kt algorithm:
\begin{align}
  I_{q;q+g}^R = (5.33)[1]
 \to I_{q;q+g}^R + \frac{\as}{2\pi}  (k^2)^{\gamma-\frac{n}{2}} (k\cdot l)^n
 \sum_{a=q,\bar{q}} f_a(x_J) C_F \left(3-\frac{\pi^2}{3}-3\log2 \right) \;.
\label{diffIqg}
\end{align}

In the {\bf gluon-initiated} case the procedure is identical; here we have two
contributions: one from the $q\bar{q}$ jet
\begin{align}
  I_{g;q+\bar{q}}^R = (5.47)[1]
  \to I_{g;q+\bar{q}}^R + \frac{\as}{2\pi}  (k^2)^{\gamma-\frac{n}{2}} (k\cdot l)^n
  \frac{C_A}{C_F} f_g(x_J) \, 2 n_f T_R \left(\frac{2}{3}\log2 -\frac{23}{36} \right) \;,
\label{diffIqqq}
\end{align}
and one from the $gg$ jet
\begin{align}
  I_{g;g+g}^R = (5.54)[1]
  \to I_{g;g+g}^R + \frac{\as}{2\pi} (k^2)^{\gamma-\frac{n}{2}} (k\cdot l)^n
  \frac{C_A}{C_F} f_g(x_J) C_A \left(
  \frac{131}{36} -\frac{\pi^2}{3} -\frac{11}{3}\log2 \right) \;.
\label{diffIggg}
\end{align}
The integrals~(\ref{IqqgFIP},\ref{IqqgKT}) and the analogous ones for the
gluon-initiated contributions, yielding
eqs.~(\ref{diffIqg}-\ref{diffIggg}), were already considered and
computed~\cite{MuVo12} in the first study of the relation between the kt and the
cone algorithm --- the latter sharing with FIP the condition of composite jet.

%================================================================================
\subsection{Final expressions of the jet vertices\label{s:final}}
%================================================================================

The result of the jet vertex for the cone and kt algoritms is reported
below, by adding to the original expressions of IP~\cite{IP} the modifications
computed in the previous subsections and here highlighted in boldface:
$\boldsymbol{\langle\cdots\rangle_C}$ for the cone and
$\boldsymbol{\langle\cdots\rangle_K}$ for the kt. The quark part is
\begin{align}
  I_q &= \frac{\as}{2\pi} (k^2)^{\gamma} \esp{\ui n\phi} \int_{x_J}^1
  \frac{\dif\zeta}{\zeta} \sum_{a=q,\bar{q}}
  f_a\left(\frac{x_J}{\zeta}\right)
  \Bigg\{
   \left[P_{qq}(\zeta)+\frac{C_A}{C_F}P_{gq}(\zeta)\right]\log\frac{k^2}{\mu_F^2} +
  \nonumber \\
 &\quad- 2 \zeta^{-2\gamma}[P_{qq}(\zeta)+P_{gq}(\zeta)]\log
  \frac{R}{\boldsymbol{\big\langle\max(\zeta,\bar\zeta)\big\rangle_C}}-\frac{\beta_0}{2}\log\frac{k^2}{\mu_R^2}\delta(1-\zeta)
  \nonumber\\
 &\quad+C_A\delta(1-\zeta)\left\{\chi^{(0)}_{n\nu}\log\frac{s_0}{k^2}+\frac{85}{18}+\frac{\pi^2}{2}
  +\frac12\left[\psi'\left(1+\gamma+\frac{n}{2}\right)-\psi'\left(\frac{n}{2}-\gamma\right)-\chi^{(0)\,2}_{n\nu}
  \right]\right\} \nonumber\\
 &\quad+(1+\zeta^2)\bigg\{C_A\left[\frac{(1+\zeta^{-2\gamma})\chi^{(0)}_{n\nu}}{2(1-\zeta)_+}
  -\zeta^{-2\gamma}\left(\frac{\log(1-\zeta)}{1-\zeta}\right)_+\right] \nonumber \\
 &\quad+\left(C_F-\frac{C_A}{2}\right)
  \left[\frac{\bar\zeta}{\zeta^2}I_2-\frac{2\log \zeta}{\bar\zeta}+2\left(\frac{\log(1-\zeta)}{1-\zeta}\right)_+
  \right]\bigg\} \nonumber \\
 &\quad + \delta(1-\zeta)\left[ C_F\left(3\log 2-\frac{\pi^2}{3}-\frac{9}{2} 
  +\boldsymbol{\Big\langle 3-\frac{\pi^2}{3}-3\log 2 \Big\rangle_K}\right)
  -\frac{10}{9} n_f T_R\right] \nonumber \\
 &\quad+C_A \zeta+C_F \bar\zeta +\frac{1+\bar\zeta^2}{\zeta}\left[C_A\frac{\bar\zeta}{\zeta}I_1+2C_A\log\frac{\bar\zeta}{\zeta}
  +C_F \zeta^{-2\gamma}(\chi^{(0)}_{n\nu}-2\log\bar\zeta)\right]\Bigg\} \;, \label{finalQ}
\end{align}
where $\gamma\equiv\ui\nu-1/2$, $\beta_0\equiv(11 C_A-4 n_f T_R)/3$, $\psi$ is
the digamma function, while the splitting functions $P_{ab}(\zeta)$ and the
special functions $I_j(n,\gamma,\zeta)$ are reported in app.~\ref{a:def}.

The gluon part is
\begin{align}
  I_g &= \frac{\as}{2\pi} (k^2)^{\gamma} \esp{\ui n\phi} \int_{x_J}^1
  \frac{\dif\zeta}{\zeta} f_g\left(\frac{x_J}{\zeta}\right) \frac{C_A}{C_F}
  \Bigg\{
   \left[P_{gg}(\zeta)+\frac{C_A}{C_F} 2 n_F P_{qg}(\zeta)\right]\log\frac{k^2}{\mu_F^2} +
  \nonumber \\
 &\quad- 2 \zeta^{-2\gamma}[P_{gg}(\zeta)+2 n_f P_{qg}(\zeta)]\log
  \frac{R}{\boldsymbol{\big\langle\max(\zeta,\bar\zeta)\big\rangle_C}}-\frac{\beta_0}{2}\log\frac{k^2}{4\mu_R^2}\delta(1-\zeta)
  \nonumber\\
 &\quad+C_A\delta(1-\zeta)\left\{\chi^{(0)}_{n\nu}\log\frac{s_0}{k^2}
  +\frac12\left[\psi'\left(1+\gamma+\frac{n}{2}\right)-\psi'\left(\frac{n}{2}-\gamma\right)-\chi^{(0)\,2}_{n\nu}
  \right] \right. \nonumber \\
 &\qquad\qquad\qquad\qquad \left. +\frac1{12}+\frac{\pi^2}{6} +\boldsymbol{\Big\langle
     \frac{131}{36} -\frac{\pi^2}{3} -\frac{11}{3}\log 2 \Big\rangle_K} \right\} \nonumber\\
 &\quad+2C_A(1-\zeta^{-2\gamma})\left[\left(\frac1{\zeta}-2+\zeta \bar\zeta\right)\log \bar\zeta
   +\frac{\log(1-\zeta)}{1-\zeta}\right] \nonumber \\
 &\quad+C_A\left[\frac1{\zeta}+\frac1{(1-\zeta)_+}-2+\zeta \bar\zeta\right]
  \left[(1+\zeta^{-2\gamma})\chi^{(0)}_{n\nu}-2\log \zeta+\frac{\bar\zeta^2}{\zeta^2} I_2\right] \nonumber \\
 &\quad+ 2 n_f T_R \left[2\frac{C_F}{C_A}\zeta \bar\zeta+(\zeta^2+\bar\zeta^2)\left(\frac{C_F}{C_A}\chi^{(0)}_{n\nu}
   +\frac{\bar\zeta}{\zeta}I_3\right) \right.\nonumber \\
 &\qquad\qquad\qquad \left. +\delta(1-\zeta)\left( -\frac1{12}
   +\boldsymbol{\Big\langle \frac{2}{3}\log 2-\frac{23}{36} \Big\rangle_K}
  \right) \right]\Bigg\} \;. \label{finalG}
\end{align}

It is apparent from the above expression that the $\log(R)$ coefficient $A$ of the
jet vertex is independent of the jet algorithm, while the constant coefficient
$B$ depends on it.

%%%%%%%%%%%%%%%%%%%%%%%%%%%%%%%%%%%%%%%%%%%%%%%%%%%%%%%%%%%%%%%%%%%%%%%%%%%%%%%%
\section{Numerical study\label{s:nc}}
%%%%%%%%%%%%%%%%%%%%%%%%%%%%%%%%%%%%%%%%%%%%%%%%%%%%%%%%%%%%%%%%%%%%%%%%%%%%%%%%

In this section we assess the quantitative difference among the jet vertices in
the three algorithms (cone, kt, FIP) that we considered, and also the
corresponding accuracy of their small-cone approximations (SCA). We shall use
the term {\em exact} in the sense of ``without SCA''.

\subsection{Jet vertices versus $\bs{R}$}

We start by evaluating the ``exact'' jet vertices in the three algorithms --- we
employ the numerical code used in ref.~\cite{CSSW10} --- for various values of
the jet radius $R$, and we compare them with their SCA. We expect the SCA to be
better, the smaller the values of $R$, and increasing discrepancy with
increasing $R$. On the other hand, the differences among different algorithms
shouldn't vanish with $R$, according to our analysis.

\begin{figure}[ht!]
  \centering
  \includegraphics[width=0.6\linewidth]{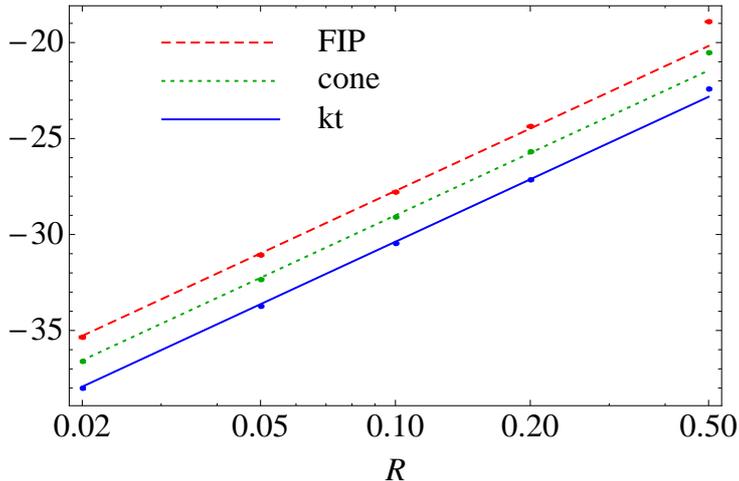}
  \caption{\it Dependence on the jet radius $R$ of the NLO jet vertices $\Psi$
    in the three jet algorithms discussed in the text: FIP (dashed red), cone
    (dotted green) and kt (solid blue). The big dots correspond to the exact
    evaluation (the small error bars showing MonteCarlo integration
    uncertainties), while the straight lines denote the small-cone
    approximation. Here $n=0$, $\nu=0$, $y=3.6$, $E=35\gev$.}
  \label{f:scaVSR}
\end{figure}

This is actually the case, as can be seen in fig.~\ref{f:scaVSR}, where we plot
the exact (points) and small-cone (lines) NLO part of the jet vertex versus $R$,
in the three algorithms mentioned before (for a given choice of the parameters
$n$, $\nu$, $E$, $y$).  The common slope of the lines represents the coefficient
$A$ of $\log(R)$, which is the same for the three algorithms, while the
intercepts at $R=1$ give the constant coefficients $B$, which clearly depend on
the algorithm.  A detailed study, carried out with several values of the
parameters, shows that the small-cone approximation works very well up to $R$ of
few tenths, with an error below 1\% for $R\lesssim0.2$ which increases up to
3-4\% when $R=0.5$.

\subsection{SCA versus algorithm choice}

Next, we specialize our analysis to realistic values of the jet radius and
energy. Since the typical phenomenological studies on MN jets use $R\simeq 0.5$
and jet transverse energies $E\gtrsim 35\gev$, in the following all quantities
will be evaluated at $R=0.5$ and $E=35\gev$. In addition, nowadays the mostly
used jet algorithm is the kt. Therefore we adopt ``exact'' quantities, computed
in the kt algorithm, as reference quantities, and we estimate the deviations to
them introduced by the SCA.

To some extent, adopting the SCA at fixed $R$ is a sort of choosing a jet
algorithm. The natural question then arises: how does the discrepancy introduced by
the small-cone approximation
\begin{equation}\label{disc1}
  (\text{exact-kt}) - (\text{SCA-kt})
\end{equation}
compare with the discrepancy caused by different choices of jet algorithm, i.e.,
\begin{equation}\label{disc2}
  (\text{SCA-kt}) - (\text{SCA-FIP}) \;?
\end{equation}

In order to answer this question, we compute the exact NLO part%
\footnote{We recall that the LO part is independent of the jet algorithm.}
of jet vertex $\Psi_{n\nu}^{\nlo}$ in the kt algorithm as function of $\nu$,
and compare it with the SCA in the same algorithm and also with the SCA in the
FIP algorithm.

\begin{figure}[ht!]
  \centering
  \includegraphics[width=0.6\linewidth]{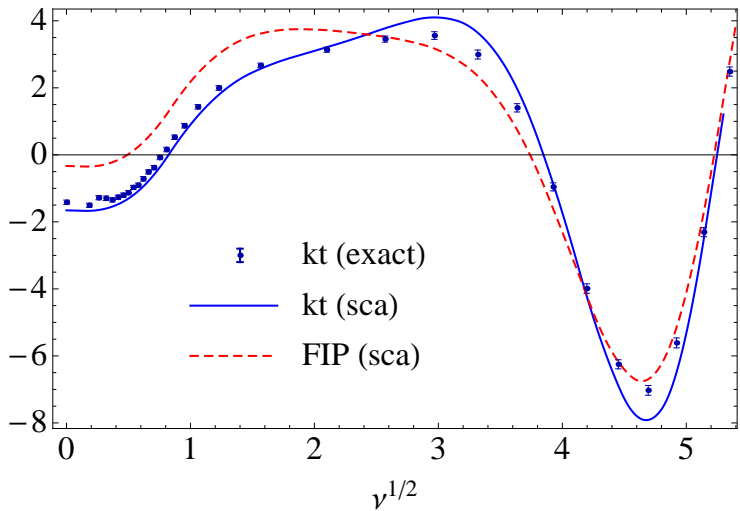}
  \caption{\it Comparison of the NLO exact jet vertex in the kt algorithm
    (points) with its SCA in the same algorithm (solid blue) and in the FIP
    algorithm (dashed red). Here $R=0.5$, $n=1$, $y=3.6$, $E=35\gev$.}
  \label{f:scaVSnu}
\end{figure}

Fig.~\ref{f:scaVSnu} shows such a comparison for $n=1$.  We can see that the
exact result in the kt algorithm is much better approximated by the SCA in the
same algorithm rather than by the FIP choice, in particular at small values of
$\nu$, which are the most important in the $\nu$ integrals of
eqs.~(\ref{Cm1},\ref{intCm}) --- since the GGF is peaked around $\nu=0$.

\begin{figure}[ht!]
  \centering
  \hspace{0.036\linewidth}\includegraphics[width=0.411\linewidth]{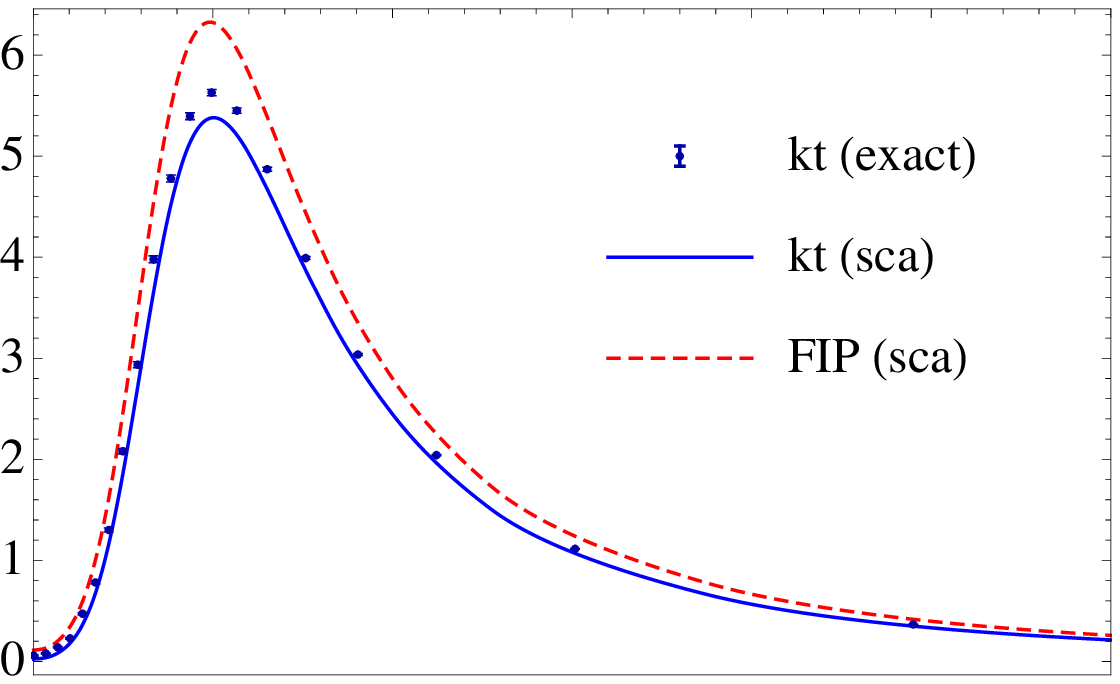}%
  \hspace{0.093\linewidth}\includegraphics[width=0.428\linewidth]{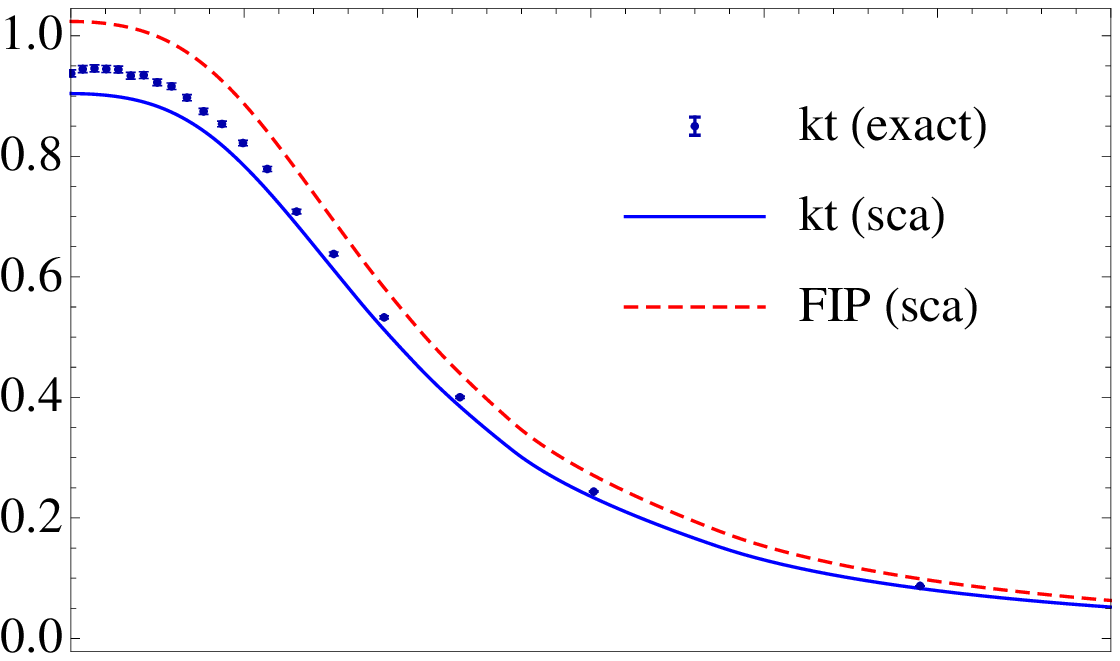}%
  \hfill\null\\
  \hfill%
  \includegraphics[width=0.44\linewidth]{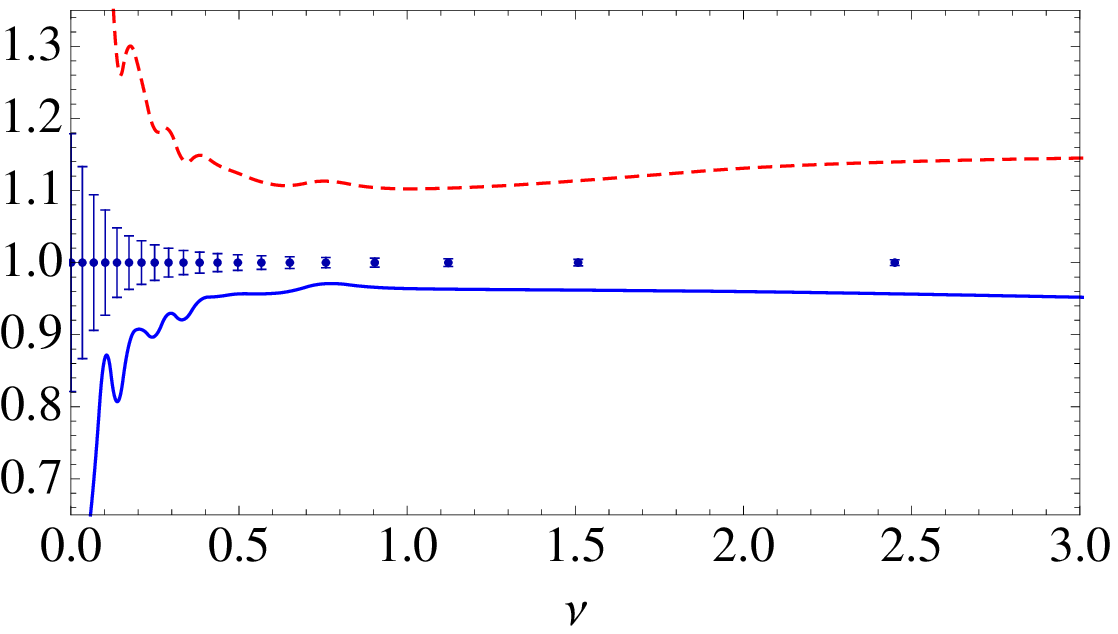}%
  \hspace{0.08\linewidth}%
  \includegraphics[width=0.44\linewidth]{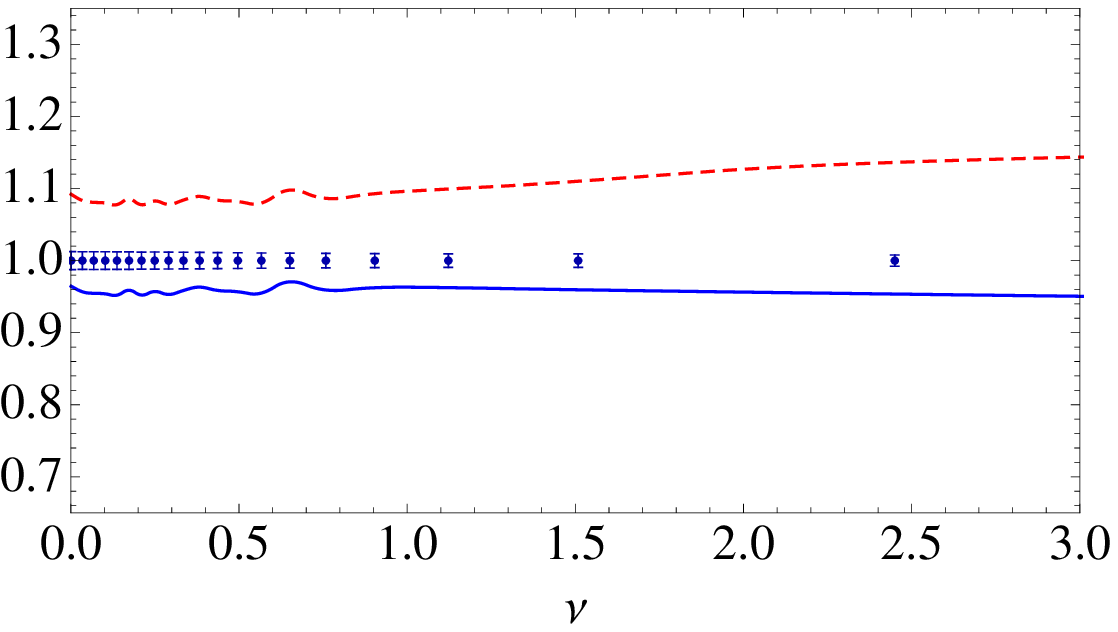}%
  \hfill\null
  \caption{\it Comparison of the exact $\nu$-integrand for kt algorithm (points)
    with the SCA in the same algorithm (solid blue) and in the FIP algorithm
    (dashed red). On the left: $n=0$; on the right: $n=1$.  The upper plots display
    absolute values in arbitrary units, and below them we show the ratios
    w.r.t.\ exact integrand. The parameters are $R=0.5$, $E_1=E_2=35\gev$,
    $y_1=3.6$, $y_2=2.8$.}
  \label{f:integrandVSnu}
\end{figure}

This conclusion is further supported by analysing the whole (LO+NLO)
$\nu$-integrand of eq.~(\ref{intCm}), again by comparing exact-kt, SCA-kt and
SCA-FIP, as in fig.~\ref{f:integrandVSnu}. The discrepancy introduced by the SCA
in the wrong FIP algorithm is about three times larger than that
introduced by the SCA in the proper kt algorithm, the latter being of the order
of 5\%.

Actually, the relative error due to the SCA is slightly larger for the full
(LO+NLO) quantities than for the pure NLO ones. This is due to the fact that the
NLO corrections usually have sign opposite to the common LO terms, giving
rise to cancellations in their sum that amplify the relative differences.

\subsection{Cross section and angular coefficients}

Finally, we present the results of the differential cross section
$\dif\sigma/\dif Y$ and of few angular coefficients $C_m/C_n$.

\begin{figure}[ht!]
  \centering
  \includegraphics[width=0.61\linewidth]{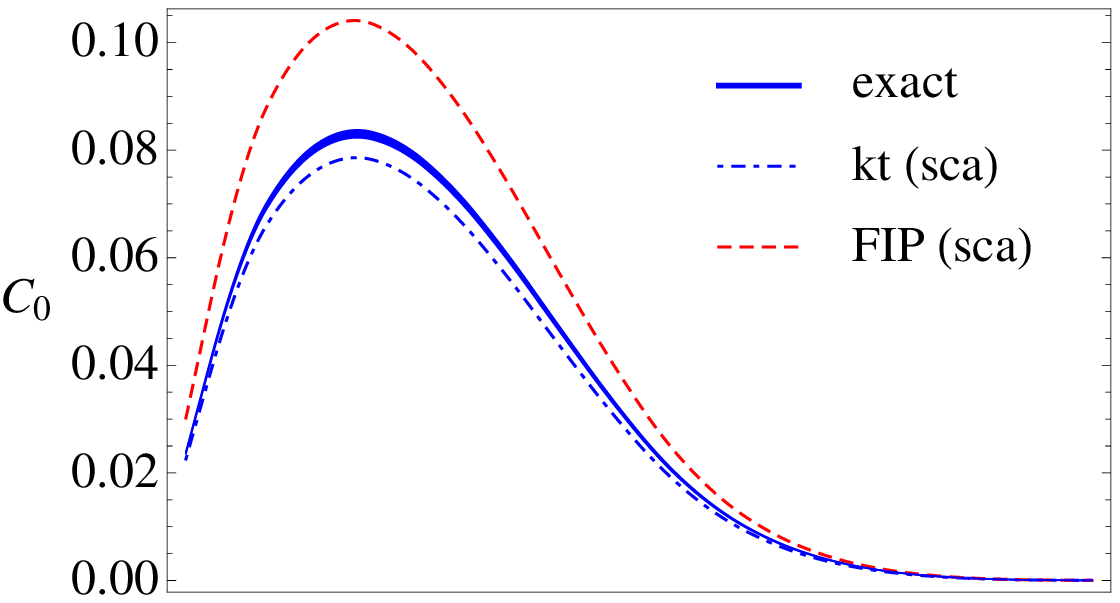}\\
  \hspace{9pt}\includegraphics[width=0.6\linewidth]{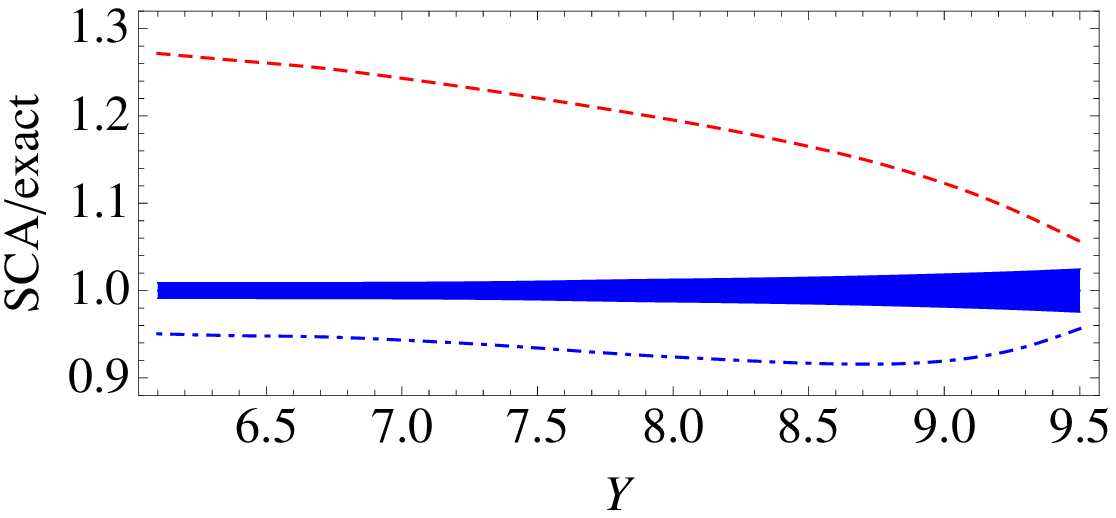}
  \caption{\it Comparison of the exact differential cross section
    $\dif\sigma/\dif Y$ ($n=0$) for kt algorithm (solid blue) with the SCA in
    the same algorithm (dash-dotted blue) and in the FIP algorithm (dashed red).
    Top: absolute values in linear scale; bottom: ratios of the SCAs w.r.t.\ the
    exact one. Here $R=0.5$, $E_1=E_2=35\gev$.}
  \label{f:dsigmadY}
\end{figure}

In fig.~\ref{f:dsigmadY} we plot the differential cross section
$\dif\sigma/\dif Y = C_0(Y)$ by comparing again the exact kt calculation with
the small-cone approximations in the KT and FIP algorithms.

It is evident that the wrong choice of the algorithm yields a large error,
especially at lower values of $Y$, while the sole SCA with the proper algorithm
introduces an error of 4-8\%.

The shape of the $C_0$ curves, which are not monothonically decreasing
in $Y$ as one could naively expect, is due to an additional cut in
rapidity $|y_i|>y_{\min}=3$ that we have imposed just for computational
convenience, as will be shortly explained. Due to this cut, the
minimum value of $Y$ that we allow is $Y_{\min} = 2 y_{\min} = 6$ and in
this limit the cross-section vanishes. It then quickly rises for $Y>6$
before eventually decreasing at larger $Y\gtrsim 7$.

The reason for imposing the $|y_i|>y_{\min}$ cut is due to the fact that
monochromatic (fixed $E$) impact factors oscillate at large $\nu$, causing the
$\nu$ integration in eq.~(\ref{Cm1}) to be slowly convergent.  Actually, the
convergence is provided by the GGF, whose modulus decreases at large $\nu$, the
decrease being faster at larger values of $Y$. On the other hand, at low values
of $Y$, and in particular with very asymmetric rapidity configurations
($\big||y_1|-|y_2|\big|\gg 1$), the numeric $\nu$-integration has to be pushed
to large $\nu$-values, thus demanding a large computational effort. Since this
problem is absent for realistic phenomenological studies (where the impact
factors are integrated in the $E$ variable and decrease themselves with
$\nu$), and because the goal of this analysis is just to compare the main
features of the different algorithms, we solve the convergence issue by imposing
the mentioned cut in rapidity $|y_i|>3$.

\begin{figure}[th!]
  {\centering
  \hspace{0.03\linewidth}\includegraphics[width=0.45\linewidth]{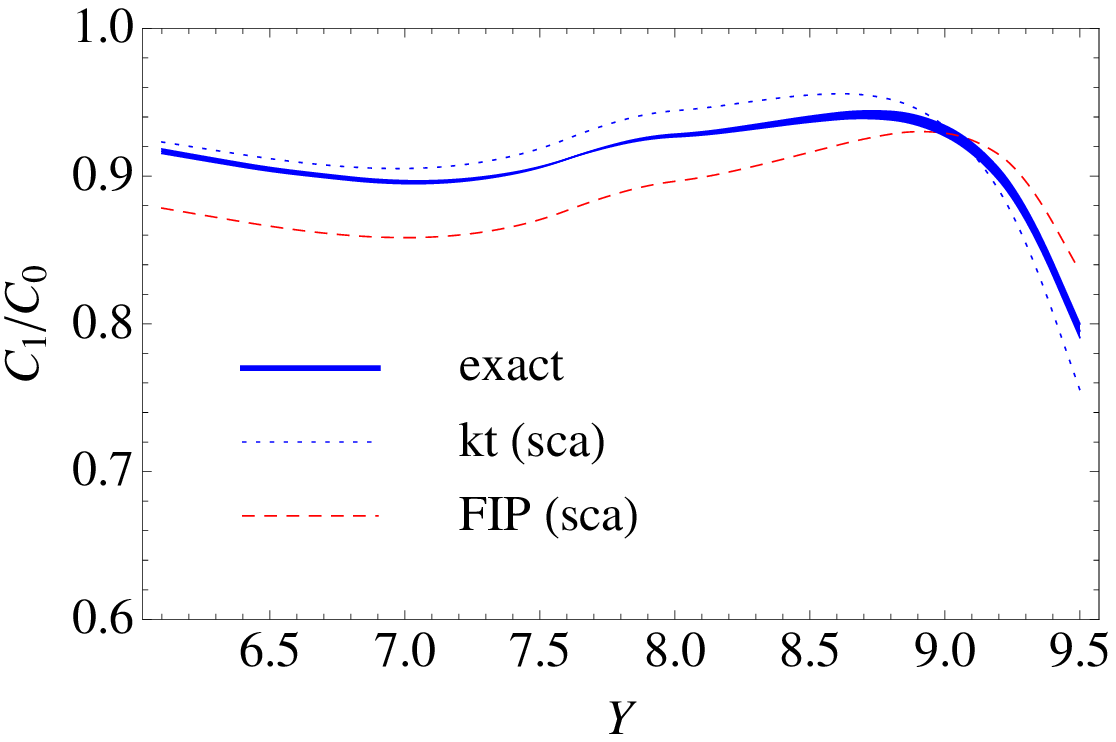}\hfill
  \includegraphics[width=0.46\linewidth]{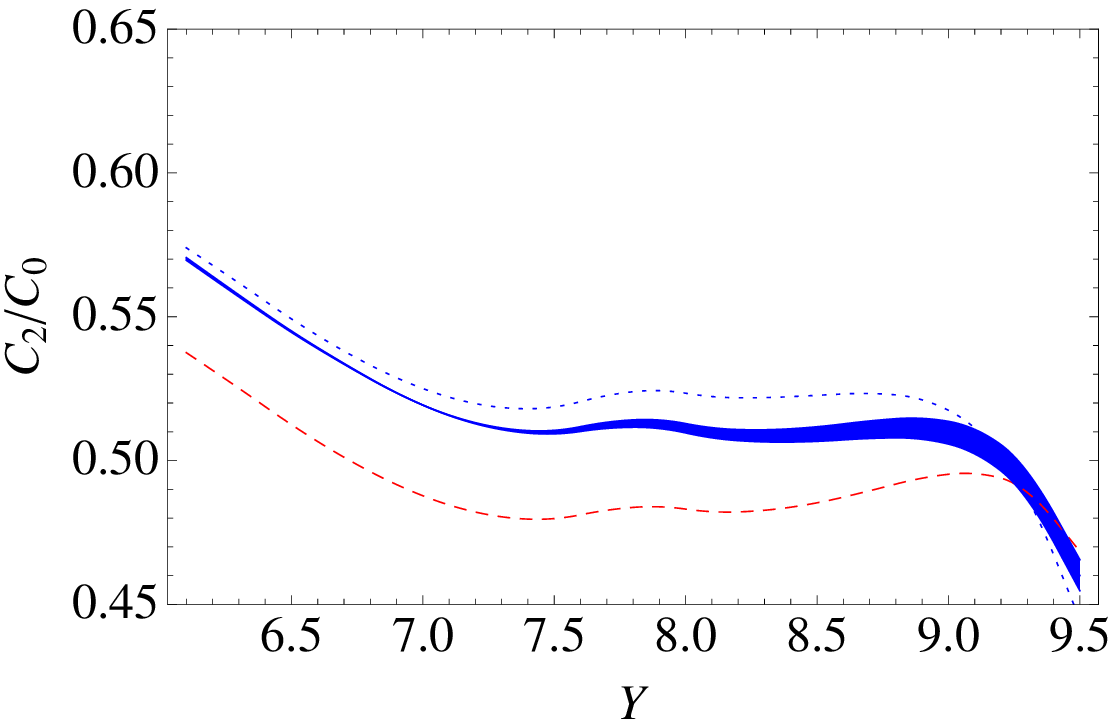}\\}
  \hspace{0.02\linewidth}\includegraphics[width=0.46\linewidth]{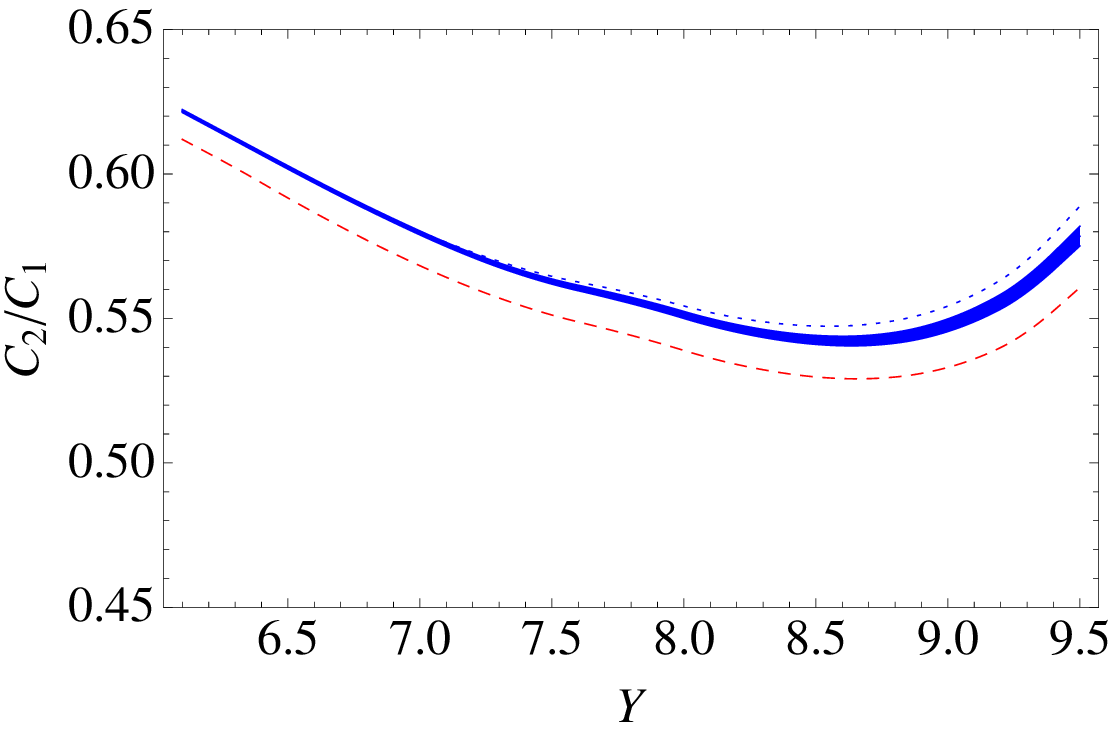}
  \caption{\it Comparison of some ratios $C_m/C_n$ determined with the exact jet
    vertices in the kt algorithm (solid blue) with those obtained in the SCA
    with the same algorithm (dotted blue) and with the FIP algorithm (dashed
    red). Here $R=0.5$, $E_1=E_2=35\gev$.}
  \label{f:CmCn}
\end{figure}

From the plots of figs.~\ref{f:integrandVSnu},\ref{f:dsigmadY} one can infer
that the main effect of the SCA is mostly an overall normalization change, and
that this is also true even for the wrong choice of algorithm (kt versus FIP),
though with a larger factor. If this were the case, by computing {\em ratios} of
observables such effects should cancel out and reproduce more faithfully the
exact quantities. This is partially true, as can be seen in fig.~\ref{f:CmCn},
where we plot some ratios of angular coefficients
$C_m(Y)/C_n(Y)=\langle\cos(m\Delta\phi_J)\rangle/\langle\cos(n\Delta\phi_J)\rangle$
which are usually adopted in order to measure the azimuthal decorrelation of the
MN jets.

It is nevertheless evident that the cancellation of the systematic effects is
more effective if the SCA is made with the proper algorithm, leading to a
discrepancy of about 2\% or less for all the ratios considered. A different
choice of algorithm yields definitely larger discrepancies, and therefore should
be avoided.

%%%%%%%%%%%%%%%%%%%%%%%%%%%%%%%%%%%%%%%%%%%%%%%%%%%%%%%%%%%%%%%%%%%%%%%%%%%%%%%%
\section{Conclusions\label{s:conc}}
%%%%%%%%%%%%%%%%%%%%%%%%%%%%%%%%%%%%%%%%%%%%%%%%%%%%%%%%%%%%%%%%%%%%%%%%%%%%%%%%

In this paper we have reconsidered the high-energy factorization formula for
Mueller-Navelet jet production, and we have computed the NLO jet vertices in the
small-cone approximation for two different jet algorithms: the {\em kt algorithm}
and the {\em cone algorithm}.

The small-cone approximation amounts to evaluate the generic jet vertex
$\Psi$ at small jet radius $R\to 0$ according to the expansion
$\Psi=A\log(R)+B+\ord{R^2}$, where the coefficient $A$ is universal, depending
only on the infra-red properties of QCD, while the constant term $B$ depends on
the jet algorithm. It should be noted that the neglected term in the expansion
is quadratic in $R$, and that the linear term is missing. It turns out that the
small-cone approximation is quite accurate for realistic values of $R\lesssim
0.5$ and, because of its analytic and computational simplicity, it has been
and will be used with benefit.

The calculation of the Mellin-Fourier components of the jet vertices (cfr.\
sec.~\ref{s:rms}) was originally performed~\cite{IP} by adopting an algorithm
which is equivalent to Furman's one~\cite{Fur82}, but which is not used anymore
in modern phenomenological studies, being infra-red unsafe. In order to use such
vertices for the analysis of data from high-energy colliders, we have computed
them in the two most popular algorithms: the kt and the cone ones. The ensuing
analytical expressions have been taken out in sec.~\ref{s:scjv}, and the
differences among the three algorithms have been highlighted.

The quantitative difference between the small-cone approximations in the kt
algorithm and in Furman's one have been estimated in sec.~\ref{s:nc}, by
plotting several quantities of interest: the jet vertices, the integrand of the
high-energy factorization formula, the differential cross section w.r.t.\ the
rapidity distance $Y\equiv|y_1-y_2|$ of the two MN jets, and some angular
coefficients indicating the azimuthal decorrelations of the jets
themselves. Also the corresponding quantities with the exact jet vertices in the
kt algorithm have been plotted.

It turns out that the difference between the kt and Furman's algorithms is
sizeable, of the order of 20\% at the level of cross section, and about 5\% for
the ratios $C_m/C_0=\langle\cos(m\Delta\phi_J)\rangle$ and $C_m/C_n$, the
angular coefficients of azimuthal decorrelation.

On the other hand, the discrepancies between the exact results and the
small-cone approximated ones are much smaller, of the order of 5\% at the level
of cross section, and less than 2\% for the angular ratios. We therefore
conclude that the small-cone expansion, computed with the proper jet algorithm,
provides a good approximation to the Mueller-Navelet jet vertices; it can then
be used as a very convenient tool to perform phenomenological studies, in that
it requires much less computational resources than the exact computation.

This aspect could be essential when analysing observables obtained by
integrating the jet energies in a non-factorized domain, e.g., by requiring
$E_1+E_2>2 E_{\cut}$. Such condition is often used in dijet analysis since, at
fixed NLO, it yields more stable results than the condition $E_i>E_{\min}$,
while retaining the $1\leftrightarrow 2$ symmetry. On the other hand, a
non-factorizable domain of integration in energy prevents the use of
eq.~(\ref{intCm}) with independently integrated impact factors $\intI$; a
numerical integration in energy (and possibly in rapidity) of the
integrand~(\ref{Cm1}) is thus necessary, and the small-cone approximation
represents a valuable tool to reduce the computational effort of such
calculation.

%%%%%%%%%%%%%%%%%%%%%%%%%%%%%%%%%%%%%%%%%%%%%%%%%%%%%%%%%%%%%%%%%%%%%%%%%%%%%%%%
\section*{Acknowledgements}
%%%%%%%%%%%%%%%%%%%%%%%%%%%%%%%%%%%%%%%%%%%%%%%%%%%%%%%%%%%%%%%%%%%%%%%%%%%%%%%%

We wish to thank the {\em Galileo Galilei Institute for Theoretical Physics} where
part of this work was performed.

%%%%%%%%%%%%%%%%%%%%%%%%%%%%%%%%%%%%%%%%%%%%%%%%%%%%%%%%%%%%%%%%%%%%%%%%%%%%%%%%%
%%%%%%%%%%%%%%%%%%%%%%%%%%%%%%%%%%%%%%%%%%%%%%%%%%%%%%%%%%%%%%%%%%%%%%%%%%%%%%%%%
\appendix
%%%%%%%%%%%%%%%%%%%%%%%%%%%%%%%%%%%%%%%%%%%%%%%%%%%%%%%%%%%%%%%%%%%%%%%%%%%%%%%%%
%%%%%%%%%%%%%%%%%%%%%%%%%%%%%%%%%%%%%%%%%%%%%%%%%%%%%%%%%%%%%%%%%%%%%%%%%%%%%%%%%

%%%%%%%%%%%%%%%%%%%%%%%%%%%%%%%%%%%%%%%%%%%%%%%%%%%%%%%%%%%%%%%%%%%%%%%%%%%%%%%%
\section{Expressions of splitting and special functions\label{a:def}}
%%%%%%%%%%%%%%%%%%%%%%%%%%%%%%%%%%%%%%%%%%%%%%%%%%%%%%%%%%%%%%%%%%%%%%%%%%%%%%%%

The splitting functions found in the jet vertices are defined in the usual way:
\begin{subequations}\label{splitFun}
  \begin{align}
    P_{qq}(z) &= C_F\left( \frac{1+z^2}{1-z} \right)_+
    = C_F\left[ \frac{1+z^2}{(1-z)_+} +\frac32\delta(1-z)\right] \\
    P_{gq}(z) &= C_F\frac{1+(1-z)^2}{z} \\
    P_{qg}(z) &= T_R\left[z^2+(1-z)^2\right] \\
    P_{gg}(z) &= 2C_A\left[\frac{1}{(1-z)_+} +\frac{1}{z} -2+z(1-z)\right] +
    \left(\frac{11}{6}C_A-\frac{n_f}{3}\right)\delta(1-z)\;.
  \end{align}
\end{subequations}
The functions $I_j$ are expressed in terms of hypergeometric functions
and read~\cite{IP}
\begin{subequations}\label{Ij}
  \begin{align}
    I_2(n,\gamma,\zeta) &=\frac{\zeta^2}{\bar\zeta^2} \bigg[ \zeta
    \left(\frac{{}_2F_1(1,1+\gamma-\frac{n}{2},2+\gamma-\frac{n}{2},\zeta)}{\frac{n}{2}-\gamma-1}
      -\frac{{}_2F_1(1,1+\gamma+\frac{n}{2},2+\gamma+\frac{n}{2},\zeta)}{\frac{n}{2}+\gamma+1}
    \right) \nonumber \\
    &\quad +\zeta^{-2\gamma} \left(
      \frac{{}_2F_1(1,-\gamma-\frac{n}{2},1-\gamma-\frac{n}{2},\zeta)}{\frac{n}{2}+\gamma}-
      \frac{{}_2F_1(1,-\gamma+\frac{n}{2},1-\gamma+\frac{n}{2},\zeta)}{\frac{n}{2}-\gamma}
    \right) \nonumber \\
    &\quad
    +\left(1+\zeta^{-2\gamma}\right)\left(\chi^{(0)}_{n\nu}-2\log\bar\zeta
    \right)
    +2\log\zeta \bigg] \label{I2} \\
    I_{1,3}(n,\gamma,\zeta) &= \frac{\bar\zeta}{2\zeta} I_2(n,\gamma,\zeta)
    \pm\frac{\zeta}{\bar\zeta}
    \left[\log\zeta+\frac{1-\zeta^{-2\gamma}}{2}\left(\chi^{(0)}_{n\nu}-2\log\bar\zeta\right)\right]
    \label{I13}
%    \label{I1} \\
%    I_3(n,\gamma,\zeta) &= \frac{\bar\zeta}{2\zeta} I_2(n,\gamma,\zeta)
%    -\frac{\zeta}{\bar\zeta}
%    \left[\log\zeta+\frac{1-\zeta^{-2\gamma}}{2}\left(\chi^{(0)}_{n\nu}-2\log\bar\zeta\right)\right]
%    \label{I3}
  \end{align}
\end{subequations}
and we recall the definition $\gamma\equiv\ui\nu-1/2$. $\chi^{(0)}_{n\nu}$ is
defined in eq.~(\ref{chi0}).

%%%%%%%%%%%%%%%%%%%%%%%%%%%%%%%%%%%%%%%%%%%%%%%%%%%%%%%%%%%%%%%%%%%%%%%%%%%%%%%%%

\end{document}